\documentclass[superscriptaddress,aps,preprintnumbers,amsmath,amssymb,prd,nofootinbib,preprint,longbibliography]{revtex4-1}
\pdfoutput=1

\usepackage{epsfig}
\usepackage{amsmath}
\usepackage{amssymb}
\usepackage{slashed}
\usepackage{verbatim}
\usepackage{bm}
\usepackage{dsfont}
\usepackage{cancel}
\usepackage{hyperref}
\usepackage{xcolor}
\usepackage{hyperref}
\usepackage[normalem]{ulem}
\hypersetup{
colorlinks = true,
linkcolor=blue,
citecolor=green!40!black,
urlcolor=green!40!black,
}

\usepackage{multirow}
\usepackage{bbold}

\def\Tin{T_{\rm ent}}
\def\Tm{T_{\rm m}}
\def\msol{M_{\rm solar}}

\usepackage{soul}

\begin{document}

\title{
More axions from diluted domain walls
}
\author{Keisuke Harigaya}
\affiliation{Department of Physics, University of Chicago, Chicago, IL 60637, USA}
\affiliation{Enrico Fermi Institute, University of Chicago, Chicago, IL 60637, USA}
\affiliation{Kavli Institute for Cosmological Physics, University of Chicago, Chicago, IL 60637, USA}
\affiliation{Kavli Institute for the Physics and Mathematics of the Universe (WPI),
The University of Tokyo Institutes for Advanced Study,
The University of Tokyo, Kashiwa, Chiba 277-8583, Japan}
\author{Lian-Tao Wang}
\affiliation{Department of Physics, University of Chicago, Chicago, IL 60637, USA}
\affiliation{Enrico Fermi Institute, University of Chicago, Chicago, IL 60637, USA}
\affiliation{Kavli Institute for Cosmological Physics, University of Chicago, Chicago, IL 60637, USA}

\vskip 1cm

\begin{abstract}
We consider the scenario in which the Peccei-Quinn symmetry breaking is followed by a period of inflation. A particularly interesting case is that the string-domain wall network produced by the symmetry breaking enters the horizon after the QCD phase transition. We show that the abundance of axions produced by such a string-domain wall network is counterintuitively much larger than the conventional post-inflationary Peccei-Quinn symmetry breaking scenario. As a result, a scenario with the axion decay constant even as low as the astrophysical bound of about $10^8$~GeV can explain the observed abundance of dark matter. The axion mini-halos produced from the string-domain wall network is much more massive than the conventional scenario. We also briefly discuss models which can realize this scenario such as a Peccei-Quinn phase transition during inflation or a second inflation after a Peccei-Quinn phase transition.
\end{abstract}

\maketitle

\tableofcontents

\section{Introduction}

The Peccei-Quinn (PQ) mechanism~\cite{Peccei:1977hh,Peccei:1977ur} is an elegant solution to the strong CP problem. It predicts a new light scalar field called the axion~\cite{Weinberg:1977ma, Wilczek:1977pj}, which is also a good dark-matter candidate~\cite{Preskill:1982cy,Abbott:1982af,Dine:1982ah}.

There are two well-studied, conventional ways of producing the axion as dark matter in the early universe. The so-called misalignment mechanism~\cite{Preskill:1982cy,Abbott:1982af,Dine:1982ah} typically assumes that the PQ symmetry is already broken before the inflation epoch that is responsible for generating the density perturbations in the observable universe. A main constraint on this scenario is the level of the isocurvature perturbations~\cite{Linde:1984ti,Linde:1985yf,Seckel:1985tj,Lyth:1989pb,Lyth:1991ub,Turner:1990uz,Linde:1991km}. On the other hand, one can assume that the PQ phase transition happens after the inflation. In this scenario, cosmic strings will be generated. Around the QCD phase transition, a string-domain wall network is formed, which collapses subsequently to radiate axions~\cite{Davis:1986xc}. These two scenarios favor the axion decay constant of $10^{10-12}$ GeV. See~\cite{Kawasaki:2014sqa,Klaer:2017ond,Buschmann:2019icd,Hindmarsh:2019csc,Gorghetto:2020qws,Dine:2020pds,Buschmann:2021sdq} for the estimation of the axion abundance in the latter scenario.

\begin{figure}[h!]
\centering
 \includegraphics[width = 0.7 \textwidth]{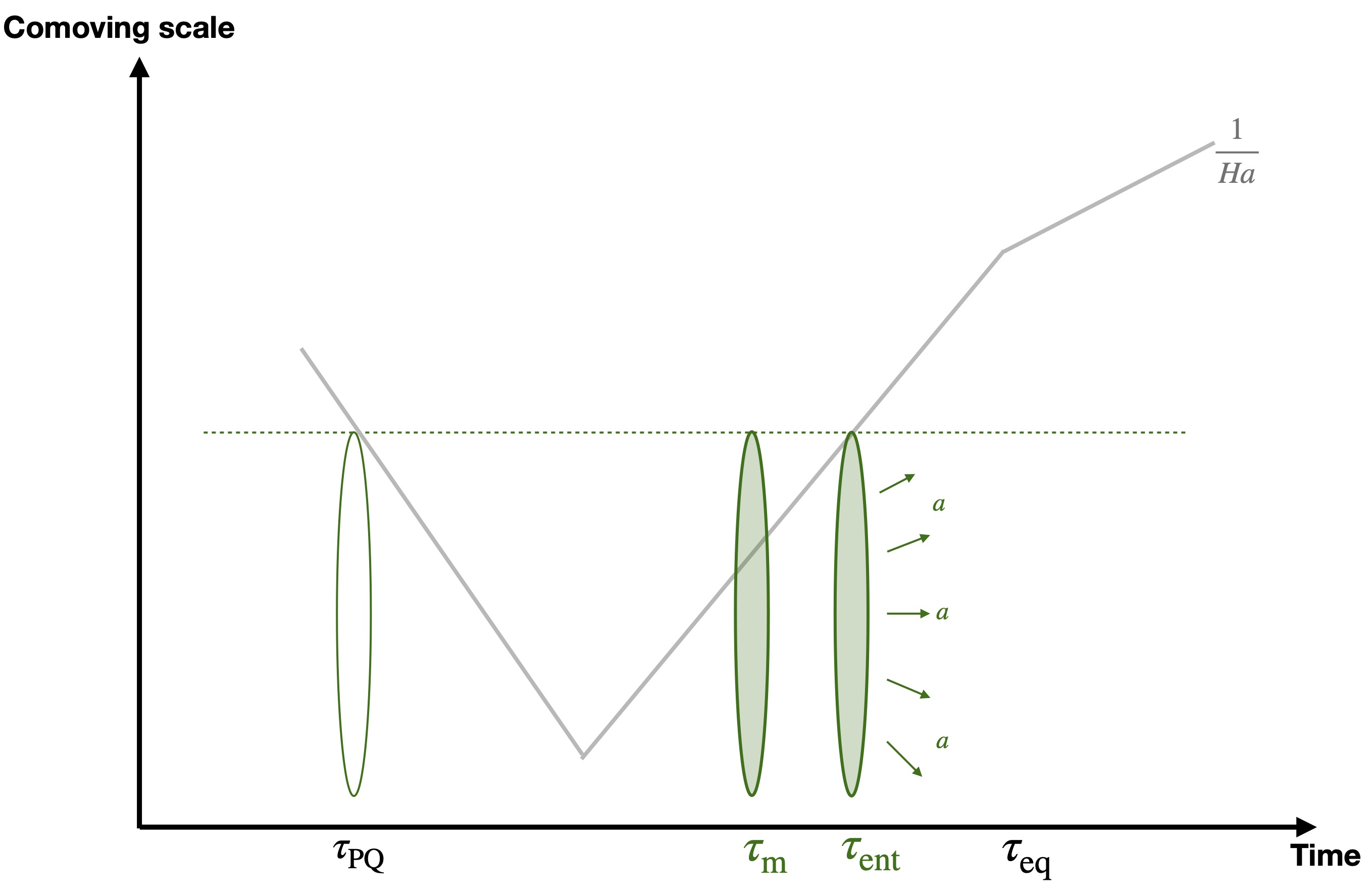} \hfill
 \caption{The simplest realization of the setup considered in this paper. The solid line shows the size of the horizon. The PQ breaking during inflationary phase, presumably after the modes corresponding to CMB and large scale structure exit the horizon. The  string network is produced after PQ symmetry breaking with the comoving size shown by the dotted line. A period of inflation occurs afterward to push network outside the horizon. After inflation ends, the axion mass becomes larger than the Hubble scale at $\tau_m$. Subsequently,  the string-domain wall network enters the horizon at $\tau_{\rm ent}$ and decays into axions. } 
 \label{fig:setup}
\end{figure}

In this paper, we discuss a different scenario in which a short period of inflation occurs after the PQ phase transition. The simplest example of this is shown in Fig.~\ref{fig:setup}, in which the PQ phase transition occurs after the epoch during which the CMB and large scale-structure fluctuation modes are generated. In this case, the axion strings will be stretched outside of the horizon by the inflation with their comoving size and separation frozen during the super horizon evolution.
After the end of inflation, the string will re-enter the horizon, i.e., the typical separations between the strings will become smaller than the horizon size. The subsequent evolution depends on when the re-entry happens relative to the QCD phase transition. If the strings re-enter significantly earlier than the QCD phase transition, it will evolve similar to the string network produced from a post-inflationary PQ phase transition.  However, there is some qualitative difference if the strings re-enter the horizon after the QCD phase transition. Around the QCD phase transition, the string-domain wall network starts to form. At the same time, since the typical separations between the strings are still larger than the horizon, causality will prevent the network from collapsing.  The network will collapse once the strings re-enter the horizon. As we will show in Sec.~\ref{sec:axion}, the subsequent production of axions are dominated by domain walls, and is enhanced in comparison with the post-inflationary PQ phase transition. As a result, this points to
a smaller decay constant than the conventional scenarios. After imposing the astrophysical constraint on the axion decay constant, $f_a > 10^8$ GeV~\cite{Ellis:1987pk,Raffelt:1987yt,Turner:1987by,Mayle:1987as,Raffelt:2006cw,Chang:2018rso,Carenza:2019pxu}, the strings can not re-enter the horizon too late without over-producing dark matter. This limits the re-entry time to be somewhat earlier than the BBN. 
This implies a coincidence; the QCD phase transition and the horizon reentry, which are dictated by completely different physics, occur at a similar time. Nevertheless, we emphasize that it is a genuine possibility that leads to qualitatively different physics. In Section~\ref{sec:model}, we briefly discuss possible ways of realizing this scenario. We also note that the inflation that stretches the strings outside the horizon does not have to be the same inflation as that generates the large scale cosmic perturbations. 

The formation of axion mini-halos~\cite{Hogan:1988mp,Kolb:1993zz,Kolb:1993hw,Kolb:1994fi} can be a sensitive probe of the axion production from topological defects. In our case, since the collapse starts later than the usual post-inflationary PQ phase transition scenario, the perturbation is generated at larger scales, leading to larger mini-halos, which can be probed by observations. In Sec.~\ref{sec:mini halos}, we estimate the properties of the mini-halos such as the mass and scale density.

\section{Axions from large domain walls}
\label{sec:axion}
In this section, we describe the evolution of strings and domain walls 
whose length scale is initially much longer than the horizon size and estimate the abundance of axions produced by the decay of them.

As a result of inflation after PQ-symmetry breaking, the typical curvature radius of cosmic strings $r_{\rm st}$, which is of the same order as the typical distance between strings, becomes much longer the horizon size $r_H = 1/H$. The displacement of strings per cosmic time $\sim 1/H$ is at the most $r_H \ll r_{\rm st} $, so the configuration of strings almost does not change and is frozen in the co-moving coordinate, $r_{\rm st}/a$.

Since $ r_{\rm st}\propto a $ and $r_H \propto a^n$ ($n=2$ for radiation domination and $n=3/2$), $r_{\rm st}$ eventually becomes comparable to $r_H$ at $\tau_{\rm ent}$, corresponding to a a temperature $\Tin$. If $\Tin$ is above the temperature at which $m_a(T) = H(T)$, which we call $\Tm$, the strings begin to follow the scaling law for $\Tm<T< \Tin$ and the scenario reduces to the standard PQ breaking after inflation.

We assume that $\Tin < \Tm$. The axion mass becomes non-negligible for $T< \Tm$ and hence domain-wall dynamics could matter. However, since the typical curvature of domain walls is also $r_{\rm st}$,  the string-domain wall network is still frozen in the co-moving coordinate so long as $T> \Tin$. At $T=\Tin$, the string-domain wall network collapses because of the domain wall tension and produces axions.

The energy density of the strings and domain walls at $T= \Tin$ is
\begin{align}
    \rho_{\rm st} \simeq & \frac{f_a^2 {\rm ln}(f_a/H) \times \pi r_{H}} {4\pi r_H^3 / 3} = \frac{3}{4} f_a^2  H^2 {\rm ln}(f_a/H),\nonumber \\
    \rho_{\rm dw} \simeq & \frac{9 m_a(\Tin) f_a^2 \times \pi r_H^2}{4\pi r_H^3 / 3} = \frac{27}{4}  f_a^2 m_a(\Tin) H.
\end{align}
Since $m_a(\Tin) \gg H $, the domain-wall energy density is much larger than that of strings. The typical momentum of axions is $H \ll m_a(\Tin)$, so the energy of each axion is simply given by $m_a(\Tin)$. The number density of axions is equal to $\rho_{\rm dw}/m_a(\Tin)$.
Normalising this by the entropy density $s$ and multiplied by the axion mass at the zero temperature, we obtain
\begin{align}
    \frac{\rho_a}{s} \simeq 0.4~{\rm eV} \frac{f_a}{10^9~{\rm GeV}} \frac{10~{\rm MeV}}{\Tin} \left( \frac{10}{g_*(\Tin)}\right)^{1/2}.
\end{align}

\begin{figure}[t]
\centering
 \includegraphics[width = 0.7 \textwidth]{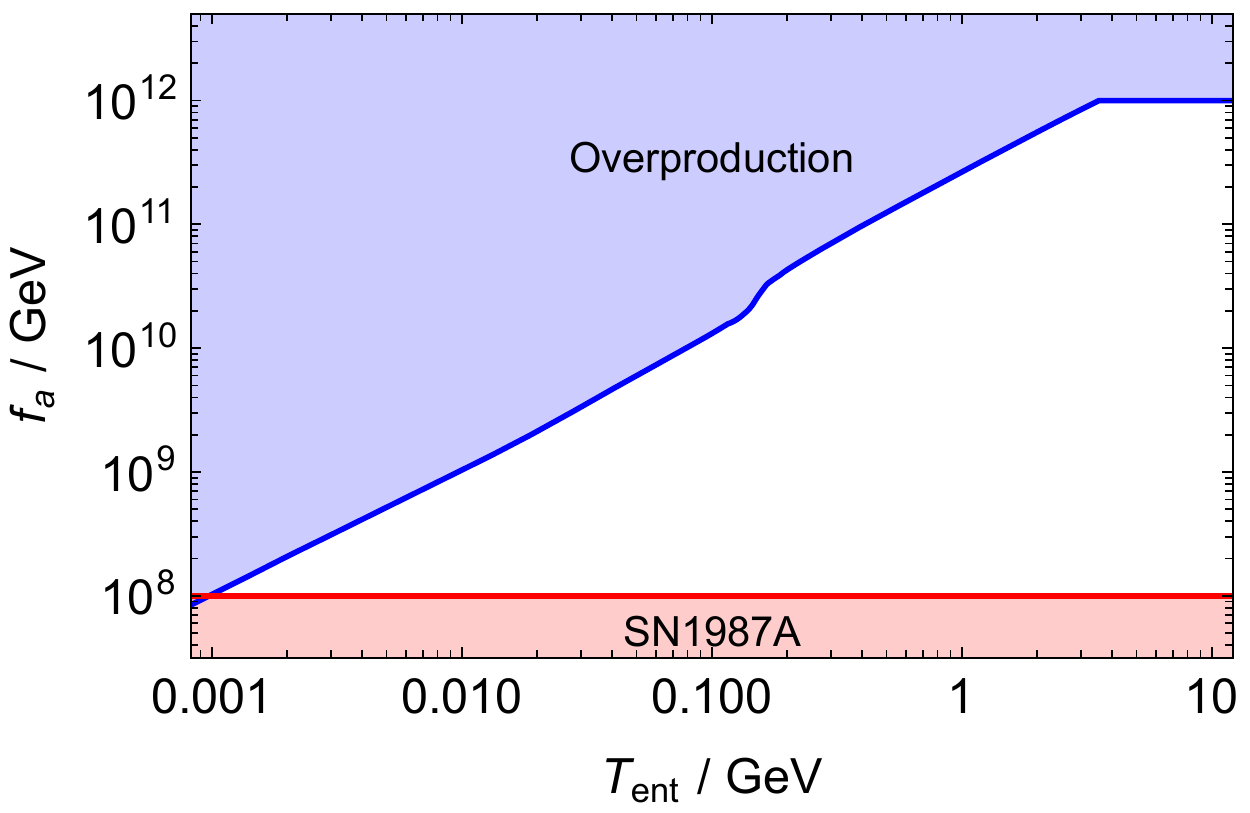} \hfill
 \caption{The required temperature  $\Tin$ at which the horizon re-entry of the string-domain wall network occurs and the axion decay constant $f_a$ is shown by the blue line. In the blue-shaded region, axion dark matter is over produced. The red-shaded region is excluded by astrophysical constraints.} 
 \label{fig:Tin}
\end{figure}

In Fig.~\ref{fig:Tin}, we show the required $(\Tin,f_a)$ to explain
the observed amount of dark matter $\rho_{\rm DM}/s \simeq 0.4$ eV by a blue line. 
One can see that $f_a$ much below the prediction of the misalignment mechanism or the post-inflationary  breaking scenario that prefer $f_a \sim 10^{10-12}$ GeV~\cite{Kawasaki:2014sqa,Klaer:2017ond,Buschmann:2019icd,Hindmarsh:2019csc,Gorghetto:2020qws,Dine:2020pds,Buschmann:2021sdq}, is consistent with axion dark matter.
For high $\Tin$ where the blue line becomes horizontal, dark matter is explained by the standard production mechanism from  strings and domain walls, where they enter the horizon when $m_a(T) < H$, follow the scaling law, and decay when $m_a(T)\sim H$. In the blue-shaded region axion dark matter is overproduced.

For $f_a$ close to the astrophysical lower bound of $10^8$ GeV~\cite{Ellis:1987pk,Raffelt:1987yt,Turner:1987by,Mayle:1987as,Raffelt:2006cw,Chang:2018rso,Carenza:2019pxu} shown by the red line, the required $\Tin$ is close to MeV. If the string-domain wall network persists even down the MeV-scale temperature, the $\theta$ term is non-zero in some part of the universe during, for example, neutron-proton conversion, and there may be some impacts on the BBN. Note that our analytical estimation of the axion abundance could be off by an $O(10)$ factor. This encourages more precise estimation of the axion abundance, perhaps via numerical lattice computation.

The enhanced axion abundance might be counter-intuitive. The string and domain walls are diluted by inflation, and still the axion abundance is enhanced. This is because of the relatively slow decrease of the energy density of domain walls and the increase of the lifetime of them by the horizon exist.
This can be explicitly seen by following the energy density of domain walls. However, it 
strongly depends on the temperature because of the dependence of the axion potential on the temperature. Instead,  it is more convenient to follow the would-be energy of domain walls which is defined as 
the surface area of regions with a misalignment of $\pi$ times the would-be tension of domain walls $\sim m_a(T=0) f_a^2$. We call this would-be domain wall energy, which becomes the actual domain wall energy density after the QCD phase transition.

Let us consider the following typical evolution.
The string-domain wall network exits the horizon at $t=t_I$. Inflation with an energy density $\rho_{\rm inf}$ lasts for an e-folding of $N_{IE}$ afterward. After the inflation, the inflaton oscillates around the minimum so the universe is matter dominated. Reheating ends at a temperature $T_R$.
The horizon-in temperature $\Tin$ can be estimated by computing the scaling of $1/(aH)$ in each era;
\begin{align}
 e^{-N_{IE}} \times \left(\frac{\rho_{\rm inf}}{T_R^4}\right)^{1/6} \times \frac{T_R}{\Tin} = 1~\longrightarrow~\Tin =  e^{-N_{IE}} \times \left(\frac{\rho_{\rm inf}}{T_R^4}\right)^{1/6} \times T_R. 
\end{align}
We denote the would-be energy fraction of the domain walls at $t_I$ as $r_I$. When the length scale of the domain walls is longer than the horizon size, the would-be energy fraction decreases during inflation in proportion to $a^{-1}$, but increases in proportion to $a^{2}$ and $a^3$ during matter and radiation domination.  The energy fraction at $\Tin$ is given by
\begin{align}
    r_I \times e^{- N_{IE}} \times \left(\frac{\rho_{\rm inf}}{T_R^4}\right)^{2/3} \times \left(\frac{T_R}{\Tin}\right)^3 = e^{2 N_{IE}} \times \left(\frac{\rho_{\rm inf}}{T_R^4}\right)^{1/6},
\end{align}
which becomes larger for lager $N_{\rm IE}$.

\section{Simple model realizations}
\label{sec:model}

In this section, we discuss how small amount of inflation can occur after PQ symmetry breaking such that topological defects can reenter the horizon much before the recombination.

\subsection{PQ symmetry breaking during inflation}

If the PQ symmetry breaking occurs during observable inflation, the amount of inflation after the PQ breaking may be small enough.  A natural starting point would be to rely on finite effective temperature corrections to the effective potential. However, this is challenging in the simplest inflation models. The effective temperature, or the Hubble scale during inflation $H_{\rm inf}$, varies slowly during the inflation.
Hence, it requires some fine tuning to arrange the PQ symmetry breaking occurring after large scale-structure modes exit the horizon and before the end of inflation. Therefore, we need to go beyond the simplest models. 

We begin with the case of $H_{\rm inf}$ less than the typical mass scale of the PQ-field potential, hence it does not play a role in triggering the phase transition. 
PQ breaking during inflation can occur if the inflaton $\phi$ directly couples to the PQ-breaking field $P$, with $\langle P \rangle = f_a/\sqrt{2}$. We consider a natural form of coupling 
\begin{align}
\label{eq:phiPcoupling}
    V(\phi,P) =  f(\phi/M)m_{\rm PQ}^2 |P|^2,
\end{align}
where $f(x)$ is a smooth function which has ${\cal O}(1)$ variation for $\Delta x \sim {\cal O}(1)$, and $M$ is a fundamental scale of the inflaton sector. If the inflaton potential is controlled by an approximate shift symmetry, $M$ may be identified with the scale where the shift symmetry is violated. 
If $f(x)$ is positive in the early stage of inflation  and becomes negative with the rolling of the inflaton, PQ-symmetry breaking can occur around the time which is appropriate to implement the scenario in this paper. To give an example, we take the typical scale in the PQ sector $m_{\rm PQ} \sim f_a \simeq 10^8$ GeV.     In this case, $H_{\rm inf} < 10^8$ GeV and $\rho_{\rm inf} < (10^{13} \ \text{GeV})^4$. We first assume that the Hubble scale is approximately constant during the inflation, so are the slow-roll parameters. The size of the density perturbation requires the first slow-roll parameter $\epsilon \sim 10^{-14}$. The excursion of the inflaton field during the last dozens of e-folds can be estimated as $\Delta \phi \sim N_e \sqrt{\epsilon} M_{\rm Planck} \sim 10^{12}$ GeV. Hence, the fundamental scale of
the inflaton sector, $M$,
would need to be $M \sim \Delta \phi \sim 10^{12}$ GeV. This is certainly a possibility. At the same time, an interesting alternative would be to assume that the inflation dynamics during the generation of the CMB and large scale structure modes is quite different from that of the later part of the inflation. In particular, the part of inflation relevant for the PQ-phase transition could have $\epsilon \sim 10^{-1} - 10^{-2}$. In this case, we would have $M \sim \Delta \phi \sim M_{\rm Planck}$ and the coupling in Eq.~(\ref{eq:phiPcoupling}) could be easily motivated by string-theory UV completions, since the inflaton shift symmetry-breaking scale is now $M_{\rm Planck}$ as expected. 

We can also consider the case in which $H_{\rm inf}$ in the early stage of inflation is somewhat higher than the desired PQ symmetry-breaking scale. We can make the same assumption that the later stage of inflation can be quite different from the CMB and large scale structure epoch. In this case, a changing $H_{\rm inf}$, through either finite effective temperature effect or couplings of $P$ with the Ricci scalar or the inflaton potential, can trigger the needed PQ phase transition.

\subsection{Second inflation after PQ-symmetry breaking}

The PQ-symmetry breaking does not have to occur during an inflationary phase. To realize our scenario, the only requirement is that there is a period of inflation after the PQ symmetry is broken. For example, 
it is possible that two inflationary phases exist after the observable universe exists the horizon. The first inflation generates the density perturbations imprinted on the CMB and large scale structure. After the first inflation ends and the universe reheats, the PQ symmetry can be broken as the temperature of the universe drops. As the universe cools further, the second inflation occurs. This can be any types of slow-roll inflation or none slow-roll inflation such as thermal inflation~\cite{Yamamoto:1985rd,Lyth:1995ka}. Another example would be that the PQ symmetry is broken during the first inflation (not necessarily while the observable universe exits the horizon), restored after it thermally or non-thermally, broken again, and  followed by the second inflation.

\section{Mini-halos}
\label{sec:mini halos}

In this section, we discuss mini-halos produced from the fluctuations sourced by large domain walls.
The axions produced from domain walls are non-relativistic, so the axion energy density should trace domain walls. The number of domain walls per a horizon volume is around one when axions are produced, so we expect that the axion energy density has $O(1)$ fluctuations with a spacial size around the horizon length at that time. Once the universe enters the matter-dominated era, those fluctuations will collapse to form mini-halos.

Let us estimate the typical mass, scale density, scale radius, and virial velocity of the mini-halos. The collapse will not be spherical, but we use the estimation assuming a spherical collapse expecting that it captures qualitative features. Also, precise estimation requires the spectrum of the fluctuation. In particular, the scale density cubically depends on $\delta \rho_a/\rho_a$. We estimate various quantities as a function of the temperature at which the collapse occurs, $T_{\rm col}$, anticipating that $T_{\rm col}$ is close to the matter-radiation equality temperature $ T_{\rm eq}$.

The typical mass of the mini-halos $M_{\rm MH}$ is the horizon mass of dark matter at $T= \Tin$,
\begin{align}
    M_{\rm MH} \simeq \frac{4\pi}{3}\left(\frac{\rho_{\rm DM}}{s}\right) \left.\frac{s}{H^3}\right|_{T=\Tin} \simeq 6\times 10^{-8} \msol \left( \frac{100~{\rm MeV}}{\Tin}\right)^3 \left(\frac{20}{g_*(\Tin)}\right)^{1/2}.
\end{align}
The typical density inside halos, called the scale density,  is
\begin{align}
\label{eq:scale density}
    \rho_s \sim 200 \rho_{\rm col} \sim 10^5 \msol {\rm pc}^{-3} \left(\frac{T_{\rm col}}{T_{\rm eq}}\right)^3,
\end{align}
which is much larger than that of standard halos formed in the CDM model because of the early collapse.
The scale radius, inside which the majority of dark matter resides, is
\begin{align}
    r_s \sim \frac{1}{(200)^{1/3}H(\Tin)} \times \left( \frac{s(\Tin)}{s(T_{\rm col})}\right)^{1/3} \sim 10^{-5}{\rm pc} \frac{100~{\rm MeV}}{\Tin} \left(\frac{20}{g_*(\Tin)}\right)^{1/6} \frac{T_{\rm eq}}{T_{\rm col}}.
\end{align}
Finally, the virial velocity is
\begin{align}
    v_{\rm vir} \sim (200)^{1/6}\frac{\rho_{\rm col}^{1/6} M_{\rm MH}^{1/3}}{M_{\rm pl}} \sim 10^{-8} \frac{100~{\rm MeV}}{\Tin} \left(\frac{20}{g_*(\Tin)}\right)^{1/6}\left(\frac{T_{\rm col}}{T_{\rm eq}} \right)^{1/2}.
\end{align}
One can check that the de Broglie length of the axion $(m_a v_{\rm vir})^{-1}$ is much shorter than the scale radius and the wave-nature of axion is not important in determining the halo profile.

The mini-halos may be observed through their gravitational interaction in the present universe. Whether or not the mini-halos survive until now depends on if tidal disruption is likely. Since the core density of mini-halos are much larger than that of standard, larger halos, the tidal disruption during the collapse of mini-halos into larger halos is ineffective. However, mini-halos may pass through the disk, which contains stars that are much denser than mini-halos and may cause tidal disruption. We follow~\cite{Arvanitaki:2019rax} to evaluate the impact of it. When a mini-halo passes a star with an impact parameter $b$ and a relative velocity $v_{\rm rel}$, a part of the mini-halo close to the star and that far from it receive different acceleration,
\begin{align}
    \delta v \sim \frac{4 M_* r_s}{8\pi M_{\rm pl}^2 b^2 v_{\rm rel}} \sim 10^{-10} \frac{100~{\rm MeV}}{\Tin} \left(\frac{20}{g_*(\Tin)}\right)^{1/6} \frac{T_{\rm eq}}{T_{\rm col}} \left(\frac{0.007{\rm pc}}{b}\right)^2,
\end{align}
which is much smaller than $v_{\rm vir}$ as long as $T_{\rm col} \gtrsim 0.02 T_{\rm eq}$, so the mini-halo is expected to be largely unaffected even by the tidal effect from stars. Here we take $b$ to be the minimal distance between a mini-halo and a star expected while crossing the disk 100 times.

\begin{figure}[t]
\centering
 \includegraphics[width = 0.7 \textwidth]{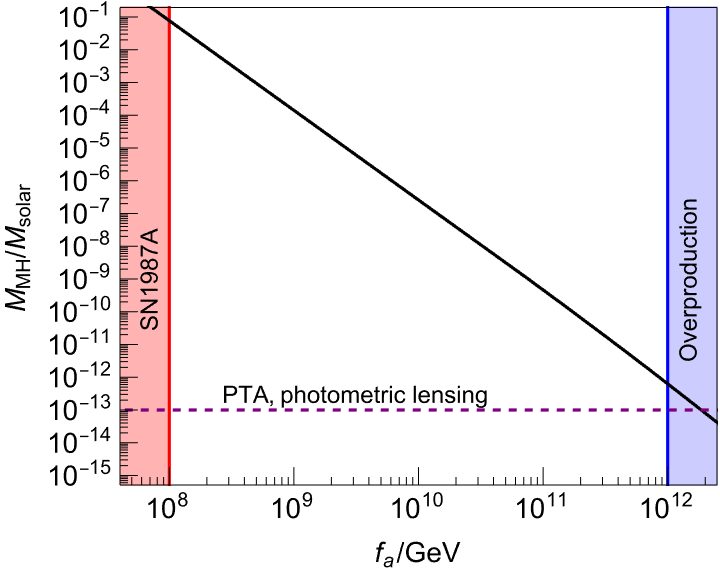} \hfill
 \caption{The typical mass of mini-halos $M_{\rm MH}$ for a given decay constant $f_a$. The sensitivity of future probe through PTA and photometric lensing (above the dashed line) is also shown.} 
 \label{fig:Mhalo}
\end{figure}

The mini-halo mass $M_{\rm MH}$ as a function of $f_a$ is shown in Fig.~\ref{fig:Mhalo}. With $\rho_s$ in Eq.~(\ref{eq:scale density}), mini-halos can be observed through pulser-timing array~\cite{Siegel:2007fz,Seto:2007kj,Dror:2019twh,Ramani:2020hdo,Lee:2020wfn} and photometric lensing~\cite{Paczynski:1985jf,Dai:2019lud,Arvanitaki:2019rax} if $M_{\rm MH} \gtrsim 10^{13} M_{\rm solar}$. The observation of mini-halos with masses consistent with the prediction in Fig.~\ref{fig:Mhalo} will be a smoking-gun signature of the scenario with large domain walls. Note the cubic dependence of the mini-halo mass on $\Tin$, which we estimated analytically. Although our estimation should capture qualitative features such as the scaling of $f_a$, it will be important to predict the axion abundance and the mini-halo mass using numerical computation.

\section{Discussion}

In this paper, we discussed a scenario where a period of inflation occurs after PQ-symmetry breaking. Somewhat counterintuitively, the abundance of axions are enhanced in comparison with the case without inflation after the PQ-symmetry breaking. The scenario leads to the formation of mini-halos whose mass is correlated with the axion decay constant. Their typical mass is much heavier than that in the conventional scenario.

We comment on other scenarios that can enhance the axion abundance and achieve axion dark matter with $f_a \ll 10^{11}$ GeV.
\begin{itemize}
    \item 
    If the radial direction of the PQ-breaking field is light, which is natural in supersymmetric theories, the radial direction can be displaced from the minimum in the early universe. The PQ-breaking field oscillates or rotates around the origin and produces axions through parametric resonance~\cite{Co:2017mop,Harigaya:2019qnl,Nakayama:2021avl} and/or the kinetic misalignment mechanism~\cite{Co:2019jts,Eroncel:2022vjg}. Our mechanism does not require a light radial direction and hence is consistent with scenarios with PQ-symmetry breaking by strong dynamics~\cite{Choi:1985cb}. The mini-halos produced by the axion fragmentation at the end of the kinetic misalignment mechanism can also produce mini-halos~\cite{Eroncel:2022efc}, but they have a different spectrum from ours. 
    \item
    If the domain-wall number is not unity, domain walls are stable and overclose the universe. We may introduce explicit PQ-symmetry breaking to let the domain walls decay into axions~\cite{Sikivie:1982qv}. By taking sufficiently small explicit breaking, we may enhance the axion abundance. This scenario predicts a non-zero strong CP phase close to the experimental upper bound~\cite{Hiramatsu:2010yn,Hiramatsu:2012sc,Kawasaki:2014sqa,Harigaya:2018ooc}, so can be distinguished from our scenario. Since the domain walls have a size of the horizon when they decay, the spectrum of mini-halos is expected to be similar to ours. 
    \item
    If the misalignment angle is close to $\pi$, the onset of the oscillation of the axion is delayed and the axion abundance is enhanced~\cite{Turner:1985si,Lyth:1991ub,Strobl:1994wk,Bae:2008ue,Visinelli:2009zm}. The large misalignment angle near the hilltop of the potential can be explained dynamically~\cite{Co:2018mho,Takahashi:2019pqf,Huang:2020etx}. Unharmonic effects near the hiltop enhances the axion perturbations and produces mini-halos~\cite{Arvanitaki:2019rax}. The spectrum of them is also different from ours.
\end{itemize}
Future detection of axion dark matter with $f_a \ll 10^{11}$ GeV means an unconventional cosmological evolution of the axion such as the ones described above. They can be distinguished from each other by the search for mini-halos and nucleon electric dipole moments.

\vspace{1cm}

\noindent
{\it Note added}: While we finalize the draft, Ref.~\cite{Redi:2022llj} appeared on arXiv, which also pointed out the enhancement of axion abundance due to domain-wall decay for the scenario of PQ-breaking during inflation. They have studied in detail the isocurvature constraints from this scenario, while our focus is on understanding of the axion parameter space that can give rise to the right relic abundance, as well as the subsequent formation of mini-halos in this scenario. The isocurvature bounds also do not directly apply to a realization of our scenario in which the PQ symmetry breaking occurs after the inflation, which is then followed by a second inflation.    

\section*{Acknowledgement}
The work of LTW is  supported by the DOE grant DE-SC-0013642.

\bibliographystyle{utphys}
\bibliography{refs}

\end{document}